\documentclass[a4paper,fleqn]{cas-dc}
\usepackage{float}
\usepackage[numbers]{natbib}
\usepackage{subfigure} 
\def\tsc#1{\csdef{#1}{\textsc{\lowercase{#1}}\xspace}}
\tsc{WGM}
\tsc{QE}
\tsc{EP}
\tsc{PMS}
\tsc{BEC}
\tsc{DE}

\begin{document}
\let\printorcid\relax
\let\WriteBookmarks\relax
\def\floatpagepagefraction{1}
\def\textpagefraction{.001}
\shorttitle{Leveraging social media news}
\shortauthors{Weiming Hu et~al.}

\title[mode = title]{
EBHI:A New Enteroscope Biopsy Histopathological H\&E Image Dataset for Image Classification Evaluation
}  

\author[a]{Weiming Hu}
\author[a]{Chen Li}
\cormark[1]
\ead{lichen201096@hotmail.com}
\author[b]{Xiaoyan Li}
\cormark[2]
\ead{lixiaoyan@cancerhosp-ln-cmu.com}
\author[c]{Md Mamunur Rahaman}
\author[b]{Yong Zhang}
\author[a]{Haoyuan Chen}
\author[a]{Wanli Liu}
\author[d]{Yudong Yao}
\author[e]{Hongzan Sun}
\author[f]{Ning Xu}
\author[g]{Xinyu Huang}
\author[g]{Marcin Grzegorzek}

\address[a]{Microscopic Image and Medical Image Analysis Group, MBIE College, Northeastern University, 110169, Shenyang, PR China}
\address[b]{Department of Pathology, Cancer Hospital, China Medical University, Liaoning Cancer Hospital and Institute,
Shenyang 110042, PR China}
\address[c]{School of Computer Science and Engineering, University of New South Wales, Sydney, NSW 2052, Australia}
\address[d]{Department of Electrical and Computer Engineering, Stevens Institute of Technology, Hoboken, NJ 07030, USA}
\address[e]{Department of Radiology, Shengjing Hospital, China Medical University, Shenyang, 110122, China}
\address[f]{School of Arts and Design, Liaoning Shihua University, Fushun 113001, China}
\address[g]{Institute of Medical Informatics, University of Luebeck, Luebeck, Germany}
\cortext[cor1]{Corresponding author}

\begin{abstract}
Background and purpose: Colorectal cancer has become the third most common cancer worldwide, accounting for approximately 10\% of cancer patients. Early detection of the disease is important for the treatment of colorectal cancer patients. Histopathological examination is the gold standard for screening colorectal cancer. 
However, the current lack of histopathological image datasets of colorectal cancer, especially enteroscope biopsies, hinders the accurate evaluation of computer-aided diagnosis techniques.\\
Methods: A new publicly available Enteroscope Biopsy Histopathological H\&E Image Dataset (EBHI) is published in this paper. To demonstrate the effectiveness of the EBHI dataset, we have utilized several machine learning,  convolutional neural networks and novel transformer-based classifiers for experimentation and evaluation, using an image with a magnification of 200$\times$.\\
Results: Experimental results show that the deep learning method performs well on the EBHI dataset. Traditional machine learning methods achieve maximum accuracy of 76.02\% and deep learning method achieves a maximum accuracy of 95.37\%.\\
Conclusion: To the best of our knowledge, EBHI is the first publicly available colorectal histopathology enteroscope biopsy dataset with four magnifications and five types of images of tumor differentiation
stages, totaling 5532 images. We believe that EBHI could attract researchers to explore new classification algorithms for the automated diagnosis of colorectal cancer, which could help physicians and patients in clinical settings.
\end{abstract}


\begin{keywords}
Enteroscope Biopsy\sep
Colorectal Histopathology\sep
Image Database\sep
Image Classification\sep
\end{keywords}

\maketitle

\section{Introduction}
Globally, colon cancer or colorectal cancer is a common and serious type of malignant cancer.
According to recent cancer statistics, colorectum is the third most common type of cancer, accounting for 10\% of all cancer cases~\cite{Sung-2021-gcs2g}.
Colon cancer and rectal cancer are often grouped together and they have many similar features when studied. In this paper, rectal, colorectal, and other types of cancers related to colon cancer are described as a broad category of colorectal cancer~\cite{pamudurthy2020advances,hossain2021machine}.
Histopathological examination of the intestine is the gold standard for the diagnosis of colorectal cancer and is a basic requirement for treatment initiation.

Enteroscope biopsy is a test that removes a small amount of biopsy tissue from the intestine for histopathological diagnosis to identify the true condition of the body, with the advantage of less damage to the body and faster healing~\cite{labianca2013early,spyrou2015comparative}.
Histopathological sections are prepared and then stained with Hematoxylin and Eosin (H\&E), a common method of staining histopathological sections to show the nucleus and cytoplasmic inclusions and to highlight the fine structure of cells and tissues~\cite{fischer2008hematoxylin,chan2014wonderful}.
When an experienced pathologist performs a histopathological examination of colorectal cancer, she or he first checks the sections for eligibility and finds the location of the lesion. Then, a low magnification microscope is used for observation and diagnosis. If the pathologist needs a clearer view of the fine structure of the lesion, the lesion can be moved to the center of the view field and switched to high magnification microscopy for further analysis~\cite{kumar2017robbins}.
However, the observation of the entire histopathological examination process leaves the following drawbacks: the diagnostic results are highly subjective and difficult to describe quantitatively; because of the workload of physicians, the section information can be easily missed out during prolonged examinations; and the diagnostic process is difficult to use big data technologies~\cite{gupta2021breast}.
As a result, there is an urgent need to address those problems more effectively.

With the advancement of computer-aided diagnosis, it is possible to examine each electronic histopathological section of colorectal cancer quickly and efficiently~\cite{mathew2021computational}. Therefore, computer vision technologies offer new possibilities to solve the problem regarding the diagnosis of colorectal cancer~\cite{pacal2020comprehensive}.

An important aspect of computer-aided analysis is image classification, the results of which can provide pathologists with important evidence in the process of histopathological diagnosis. With the development of medical image classification technology, there is an urgent need in the fields of identifying benign and malignant tumors, tumor differentiation stages, and tumor subtypes~\cite{miranda2016survey}.
To this end, we need a multi-category colorectal cancer dataset to test various medical image classification methods to obtain high classification accuracy and good robustness~\cite{kotadiya2019review}.

This paper presents a new publicly available Enteroscope Biopsy Histopathological H\&E Image Dataset (EBHI) consisting of 5532 electron microscopy images of histopathological sections of colorectal cancer, including five tumor differentiation stages, Normal, Polyp, Low-grade Intraepithelial Neoplasia (Low-grade IN), Highgrade IntraepithelialNeoplasia (High-grade IN), and Adenocarcinoma.
Four magnifications exist in this database, 40$\times$, 100$\times$, 200$\times$ and 400$\times$.
Another contribution of this paper is to classify the five types of images with 200$\times$ magnification into two major categories, Benign and Malignant. Then, five different methods are used for feature extraction and the features are classified using machine learning and deep learning-based methods.
EBHI is available at the URL: https://doi.org/10.6084/m9.figshare.16999363.v1

\section{Basic Information of EBHI}
\subsection{Database overview}

The colorectal cancer database contains 5532 histopathological images of five categories, divided into four magnifications of 40$\times$, 100$\times$, 200$\times$ and 400$\times$. 
The details of the dataset are described below.

 \textbf{
Colorectal Histopathological Image Dataset:
}
\begin{enumerate}
\itemsep=0pt
\item Data source: \\

Two pathologists from Cancer Hospital of China Medical University provide electron microscopy images of the histopathological section of colorectal cancer by enteroscope biopsy and image level annotation of the weakly supervised learning process. Four biomedical researchers organize and produce the dataset. The image annotation rules are below.\\
Rule I: When a physician finds only one differentiation stage in an image, this differentiation stage is used as the label for that image;\\
Rule II: When a doctor finds multiple differentiation stages in an image, the most significant stage is used as the label of the image;\\
Rule III: In Rule II, if different differentiation stages are similarly distributed on the image, the doctor will give the most severe stage as the label of that image.
\item Staining: H\&E staining.
\item Sampling method: Enteroscope Biopsy
\item Eyepiece Magnification: 10$\times$;\\
Objective lens Magnification: 4$\times$, 10$\times$, 20$\times$ and 40$\times$;\\
Overall magnification: 40$\times$, 100$\times$, 200$\times$ and 400$\times$.
\item Image types:\\
Normal: No cancerous cells appeared in the section (See section 2.2.1 for details);
Polyps: Augmentation lesions appeared in the section (See section 2.2.2 for details);
Low-grade IN: Mild to moderate heterogeneous hyperplasia in the section (See section 2.2.3 for details);
High-grade IN: Severe heterogeneous hyperplasia in this section (See section 2.2.3 for details);
Adenocarcinoma: Cancerous cells appear in this section (See section 2.2.4 for details).
\item Scale: See Table. \ref{tbl1}.
\begin{table}
\centering
\caption{Dataset scale of EBHI.}\label{tbl1}
\begin{tabular}{llllll}
\hline
Magnification  & 40$\times$  & 100$\times$  & 200$\times$  & 400$\times$  & Total \\ \hline
Normal         & 17  & 29   & 61   & 79   & 186   \\
Polyp          & 119 & 165  & 254  & 304  & 842   \\
Low-grade IN            & 204 & 341  & 603  & 660  & 1808  \\
High-grade IN            & 47  & 80   & 130  & 161  & 418   \\
Adenocarcinoma & 205 & 471  & 790  & 812  & 2278  \\ \hline
Total          & 592 & 1086 & 1838 & 2016 & 5532  \\ \hline
\end{tabular}
\end{table}
\item Microscope: Olympus (Japan).
\item Acquisition software: NewUsbCamera.
\item image sizes: 2048$\times$1536 pixels.
\item Image format:  "*.png".
\item All images in EBHI dataset have been withheld from all data regarding patient information. Moreover, this dataset generation work is supported by the Ethical Committee at Northeastern University, China.
\end{enumerate}

\subsection{Database description}

\subsubsection{Normal}

\begin{figure*}
	\centering
		\includegraphics[scale=.5]{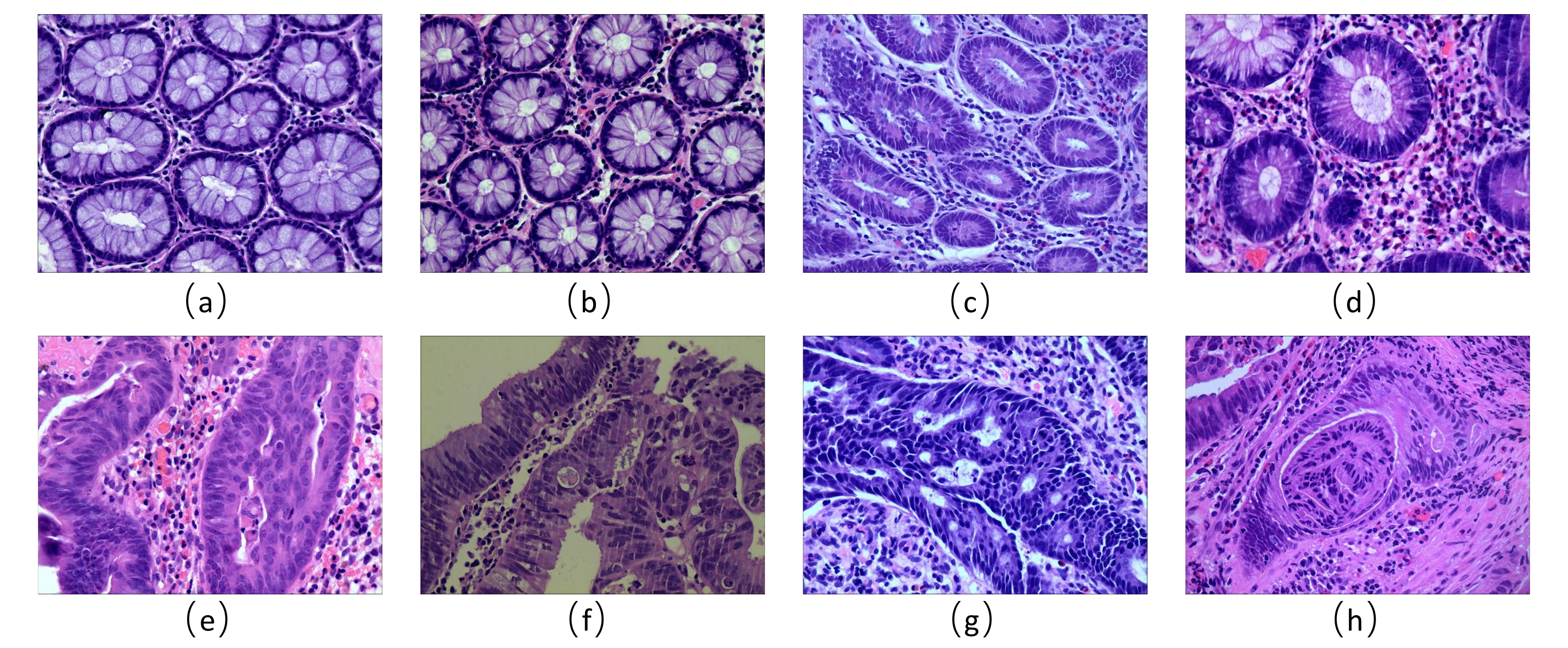}
		\caption{An example of histopathological images EBHI: (a) Normal, (b) Polyp, (c) (d) Low-grade Intraepithelial Neoplasia, (e) (f) High-grade Intraepithelial Neoplasia, (g) (h) Adenocarcinoma.}
	\label{FIG:1}
\end{figure*}

\begin{figure*}
	\centering
		\includegraphics[scale=.5]{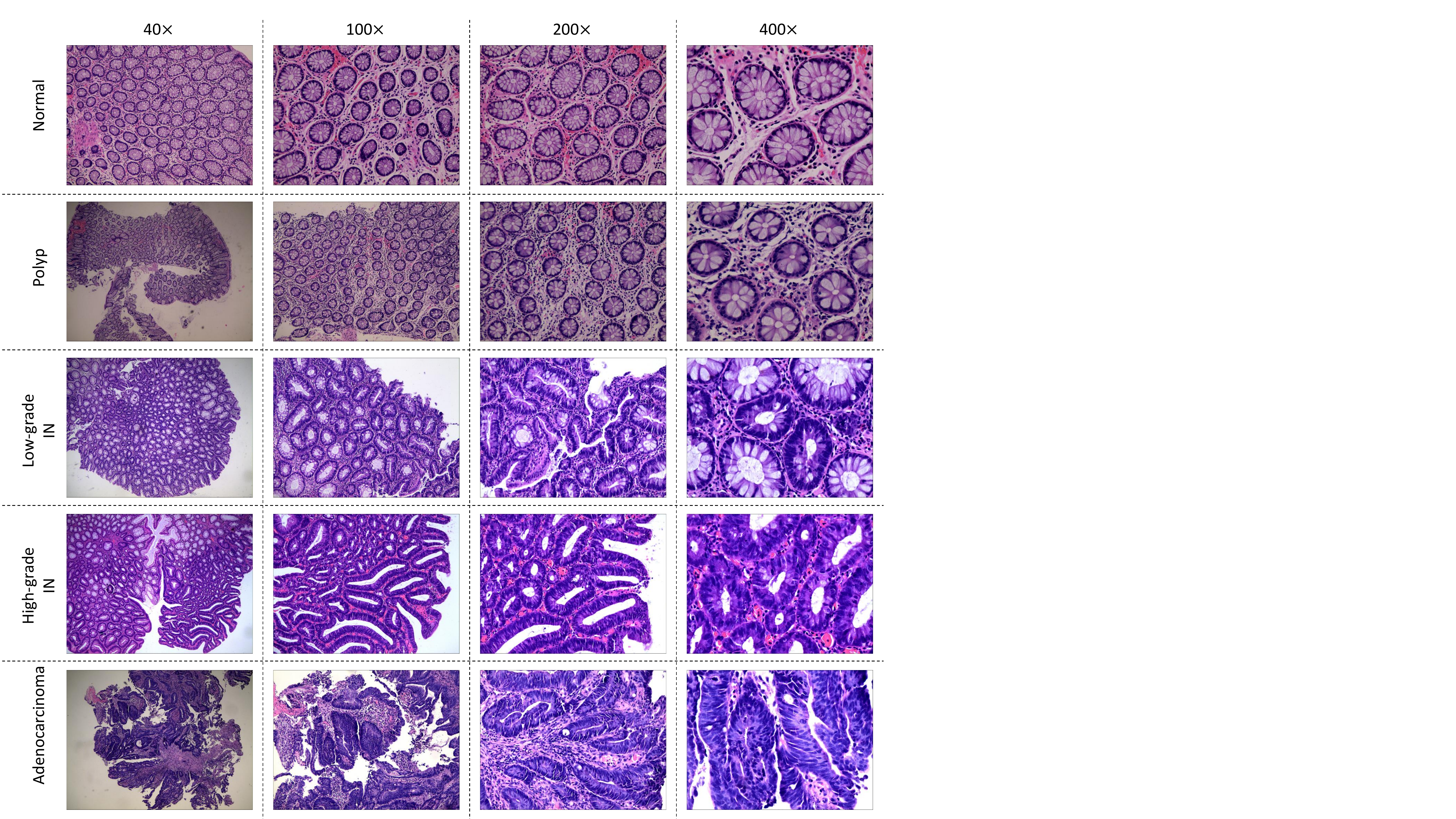}
		\caption{Sample database sections are shown, with each row in the same category and each column in the same magnification.}
	\label{FIG:2}
\end{figure*}

All areas of the image for each normal category are in the normal morphology of colorectal cells. There is no or very little heterotypic in each cell. Moreover, the nuclei of the cells in the figure are almost free of mitosis and are neatly arranged in a monolayer (Figure. \ref{FIG:1} (a)).  Therefore, when observed under an optical microscope, if no cellular and tissue lesions are observed, the sections show that the tissue has a normal tubular structure with a regular lumen arrangement. An image is considered as normal if it meets the characteristics of a normal image ~\cite{de2001pathology}.
A sample of each specific magnification is shown in the first row of Figure. \ref{FIG:2}.

\subsubsection{Polyp}

Colorectal polyps are bulging lesions on the surface of the intestinal mucosa that protrudes into the intestinal lumen and can have different histological structures~\cite{cooper1998pathology}.
The luminal structures remain intact in the histopathological section of the polyps. Nuclei division is almost absent, and the nuclei are arranged in a single layer without upward migration.
The nucleoplasmic ratio is slightly larger than in normal images but still less than 1/5 (Figure. \ref{FIG:1} (b)).
Examples of polyps from EBHI at various magnifications can be seen in the second row of Figure. \ref{FIG:2}.

\subsubsection{Intraepithelial Neoplasia}

Intraepithelial Neoplasia (IN) is a special stage prior to the development of epithelial malignancies. IN has significant changes in cell morphology and cell arrangement compared to normal tissue~\cite{ren2013missed}.
IN is divided into low grade and high grade, and the phenomenon of complex nuclei appears on IN.
There is nuclear upward migration in the IN stage and the high-grade nuclear upward migration is more severe and appears in the superficial layer.
Nuclear division is common at this stage, and the nucleoplasmic ratio increases.
The lumen structure is different and varied in shape at this stage, 
but this phenomenon is not evident at the lower levels of IN.
Figure. \ref{FIG:1} (c)(d) are histopathological examples of low-grade IN and Figure. \ref{FIG:1} (e)(f) are histopathological examples of high-grade IN.
The third row in Figure. \ref{FIG:2} shows the sample of low-grade IN at various magnifications, and the fourth row shows high-grade.

\subsubsection{Adenocarcinoma}

Adenocarcinoma is a malignant tumor in the digestive tract. Based on high-grade IN, adenocarcinoma further exhibits a dense and very irregular distribution of the lumen. Moreover, sieve-like and nest-like structures begin to appear and are accompanied by infiltrative growth~\cite{jass2012histological}. 
The borders are not clear when viewed through microscope. The nucleoplasmic ratio can exceed 1/3 at this stage.
Examples of adenocarcinoma pathology images can be seen in Figure. \ref{FIG:1} (g)(h).
The last row of Figure. \ref{FIG:2} shows the histopathological images of adenocarcinoma in EBHI at various methodological multiples.

\section{Evaluation of EBHI}
\subsection{Methods of feature extraction}

In this paper, five virtual features are extracted for image classification. They are color histogram, luminance histogram, histogram of oriented gradient (HOG), local binary pattern(LBP), and gray level co-occurrence matrix (GLCM).

\subsubsection{Color histogram and luminance histogram} 

Color and luminance histogram both are color feature extraction techniques, where the color histogram is the widely used method. It can completely describe the color distribution of the whole image, i.e., visually show the proportion of each color component in an image. At the same time, this feature is not affected by image rotation, displacement, and scaling.  It can segment images automatically that do not require the spatial location of the target. However, it is difficult to describe the local distribution of colors in an image. Luminance is a special color feature, which is the average of the three color components of a pixel point. We choose a luminance histogram to describe the luminance information.

\subsubsection{Texture features} 

Texture analysis is the extraction and analysis of the pattern of the gray-scale spatial distribution of an image. The texture is a reflection of the object structure, and analysis of texture can obtain important information about the object in the image, which is an important tool for image segmentation, feature extraction and classification. The texture is a feature that reflects the homogeneity of an image.
In this paper, HOG, LBP and GLCM are used to describe the texture of EBHI.

HOG feature is a feature descriptor used for object detection in computer vision and image processing~\cite{patel2020histogram}.
LBP is an efficient texture descriptor for measuring and extracting texture information from local images with significant advantages such as gray scale invariance and rotation invariance, 
and features are easy to compute~\cite{zeebaree2021multi}.
The recurrence of pixel gray scale in spatial location forms the texture of the image, and GLCM is a joint distribution describing the gray scale of two pixels with some spatial location relationship~\cite{kumar2020feature}.

\subsection{Classification}

 The five types of images are classified into two major categories according to the medical classification method: Benign (Normal, Polyps and Low-grade IN) and Malignant category (High-grade IN and Adenocarcinoma), and images with a magnification of 200$\times$ are selected for this part of the experiments.
After feature extraction, seven classical machine learning methods are used to classify EBHI dataset, including, Linear Regression (LR), $k$-Nearest Neighbor ($k$NN), Naive Bayesian classifier, Random Forest (RF), Linear Support Vector Machine (linear SVM), Non-linear Support Vector Machine (non-linear SVM), and Artificial Neural Network (ANN). Furthermore, four popular and novel deep learning methods are used for classification, including, VGG16, InceptionV3, ResNet50 and Vision Transformer (ViT).

The experiments are carried out using a local workstation with 32GB RAM and the windows 10 operating system. The GPU of the workstation is an 8GB NVIDIA Quadro RTX 4000.  For the traditional machine learning approach, we use the language Matlab. For the deep learning approach, we use python 3.6 programming and Pytorch version 1.7.1.

\subsubsection{Classical machine learning methods}

Machine learning methods used for classification interpret whether an image is benign or malignant by its visual features. 
LR is a method of obtaining a linear model, that uses the least-squares function to establish the relationship between one or more independent variables to predict as accurately as possible the true value of the output label~\cite{kumari2018linear}.
$k$NN is a simple and commonly used supervised learning method that finds the nearest $k$ samples to vote for the predicted outcome based on a prescribed computational distance method~\cite{gou2019generalized}. 
The naive Bayesian classifier is based on Bayesian decision theory in probability theory to classify images~\cite{yu2019novel}. 
RF is an integrated learning method among Bagging methods. RF is based on decision tree learner, which adds random attribute selection to the training process of decision tree and performs random sampling training to obtain the complete RF classification model~\cite{paul2018improved}. 
SVMs are classified as linear and nonlinear based on kernel functions~\cite{wang2018support}. 
In addition to linear classification using linear kernel functions, SVMs can also be efficiently classified as nonlinear using Radial Basis Function (RBF).
ANNs are classification algorithms that mimic the structural composition of human brain neurons and are trained by propagation algorithms~\cite{tamang2020forecasting}.

\subsubsection{Deep learning methods}

The concept of deep learning originates from the research of ANNs, where a 
multi-layer perceptron with multiple hidden layers is a deep learning 
structure. Deep learning forms a more abstract high-level representation 
attribute category or feature by combining low-level features to discover 
distributed feature representations of data~\cite{kora2021transfer}.

In 2014, the Visual Geometry Group and Google DeepMind developed a new deep 
convolutional neural network: VGG~\cite{simonyan2014very}.
VGG is a Convolutional Neural Network (CNN) improved on AlexNet. there are various forms of VGG models, the most commonly used in the image classification field is the VGG16 model. The core idea of VGG is to use three 3$\times$3 convolutional kernels instead of 7$\times$7 convolutional kernels, and two 3$\times$3 convolutional kernels instead of 5$\times$5 convolutional kernels. The main purpose of this structure is to guarantee the same perceptual field conditions, to enhance the depth of the network, and to enhance the neural network to some extent.

Google's InceptionNet debuted in the 2014 ILSVRC competition. InceptionNet has developed multiple versions, of which Inception-V3 is a more representative version in this large family~\cite{szegedy2016rethinking}. The main idea of the Inception architecture is to find out how to use dense components to approximate the optimal local sparse nodes.
Inception-V2 borrows the design idea of VGGNet by replacing the large 5$\times$5 convolution with two 3$\times$3, and also adds the well-known batch normalization to the network for the first time. Inception-V3 also introduces the decomposition of a larger two-dimensional convolution into two smaller one-dimensional convolutions.

In 2016, Kaiming He et al proposed various forms of ResNet to address the difficulty of training deep networks due to gradient disappearance~\cite{he2016deep}. 
Among the various forms of ResNet, the most commonly used one in the image classification field is ResNet50.
The ResNet50 team separately constructed a ResNet50 building block with "Shortcut Connection" and a down-sampling ResNet50 building block. A 1$\times$1 convolution operation is added to the main branch of the regional down-sampling building block~\cite{zheng2021mdcc}.

In 2020, Alexey Dosovitskiy et al. proposed the ViT
model~\cite{dosovitskiy2020image} by using the transformer, 
which is very effective in the field of natural language processing.
The algorithm has also achieved good results in the field of image classification, reducing the reliance on CNNs in the field of image classification. ViT crops the image into small blocks and provides a linear sequence of embeddings of these blocks as input to Transformer. ViT trains the model in a supervised manner for image classification.

\subsection{Evaluation of Classical Machine Learning based Classification Methods on the EBHI dataset}

In this section, seven classical classifiers are applied to five different image features to observe the classification performance of different classifiers. Table. \ref{tbl2} shows the classification results.
All classification experiments use the same parameters, in the $k$NN
method $k$ is set to 9, there are 10 trees in RF, and the nonlinear SVM uses RBF. In the ANN, a 2-layer network with 10 nodes are used in the first layer and 3 nodes are used in the second layer (500 training epoch, 0.01 learning rate, and 0.01 expected loss).

\begin{table*}
\caption{ Classification results of EBHI using different classifiers for five image features (In [\%]). Bold in the table indicates the highest value of the classification result for the same feature.}\label{tbl2}
\resizebox{\textwidth}{65mm}{
\begin{tabular}{cclrrrrrrrrrrrrr}
\hline
\multirow{2}{*}{\textbf{Freatures}}  & \multicolumn{1}{c}{\multirow{2}{*}{\textbf{Methods}}} & \multicolumn{1}{c}{\multirow{2}{*}{\textbf{Acc}}} & \multicolumn{1}{c}{\textbf{}} & \multicolumn{1}{l}{\textbf{}} & \multicolumn{4}{c}{\textbf{Malignant}}                                                                                                                           & \multicolumn{1}{c}{\textbf{}} & \multicolumn{1}{l}{\textbf{}} & \multicolumn{4}{c}{\textbf{Benign}}                                                                                                                             \\ \cline{6-9} \cline{12-15} 
                                     & \multicolumn{1}{c}{}                                  & \multicolumn{1}{c}{}                              & \multicolumn{1}{c}{\textbf{}} & \multicolumn{1}{l}{\textbf{}} & \multicolumn{1}{l}{\textbf{Precision}} & \multicolumn{1}{l}{\textbf{Recall}} & \multicolumn{1}{l}{\textbf{Specificity}} & \multicolumn{1}{l}{\textbf{F1-score}} & \multicolumn{1}{l}{\textbf{}} & \multicolumn{1}{l}{\textbf{}} & \multicolumn{1}{l}{\textbf{Precision}} & \multicolumn{1}{l}{\textbf{Recall}} & \multicolumn{1}{l}{\textbf{Specificity}} & \multicolumn{1}{l}{\textbf{F1-score}} \\ \hline
\multirow{7}{*}{Color histogram}     & LR                                                    & 51.23                                             &                               &                               & 51.41                                  & 49.46                               & 53.01                                    & 50.42                                 &                               &                               & 51.05                                  & 53.01                               & 49.46                                    & 52.01                                 \\
                                     & $k$NN                                                   & 68.94                                             &                               &                               & 70.83                                  & 64.67                               & 73.22                                    & 67.61                                 &                               &                               & 67.34                                  & 73.22                               & 64.67                                    & 70.16                                 \\
                                     & naive Bayesian                                        & 59.67                                             &                               &                               & 69.57                                  & 34.78                               & 84.70                                    & 46.38                                 &                               &                               & 56.36                                  & 84.70                               & 34.78                                    & 67.69                                 \\
                                     & RF                                                    & \textbf{73.57}                                    &                               &                               & 69.68                                  & 83.70                               & 63.39                                    & 76.05                                 &                               &                               & 79.45                                  & 63.39                               & 83.70                                    & 70.52                                 \\
                                     & linear SVM                                            & 55.04                                             &                               &                               & 59.22                                  & 33.15                               & 77.05                                    & 42.51                                 &                               &                               & 53.41                                  & 77.05                               & 33.15                                    & 63.09                                 \\
                                     & non-linear SVM                                        & 50.14                                             &                               &                               & 50.14                                  & 100.00                              & 0.00                                     & 66.79                                 &                               &                               & 50.14                                  & 100.00                              & 0.00                                     & 66.79                                 \\
                                     & ANN                                                   & 70.03                                             &                               &                               & 70.11                                  & 70.11                               & 69.95                                    & 70.11                                 &                               &                               & 69.95                                  & 69.95                               & 70.11                                    & 69.95                                 \\ \hline
\multirow{7}{*}{Luminance histogram} & LR                                                    & 57.22                                             &                               &                               & 57.54                                  & 55.98                               & 58.47                                    & 56.75                                 &                               &                               & 56.91                                  & 58.47                               & 55.98                                    & 57.68                                 \\
                                     & $k$NN                                                   & 72.21                                             &                               &                               & 74.70                                  & 67.39                               & 77.05                                    & 70.86                                 &                               &                               & 70.15                                  & 77.05                               & 67.39                                    & 73.44                                 \\
                                     & naive Bayesian                                        & 57.49                                             &                               &                               & 62.07                                  & 39.13                               & 75.96                                    & 48.00                                 &                               &                               & 55.38                                  & 75.96                               & 39.13                                    & 64.06                                 \\
                                     & RF                                                    & \textbf{72.75}                                    &                               &                               & 68.26                                  & 85.33                               & 60.11                                    & 75.85                                 &                               &                               & 80.29                                  & 60.11                               & 85.33                                    & 68.75                                 \\
                                     & linear SVM                                            & 53.41                                             &                               &                               & 52.79                                  & 66.85                               & 39.89                                    & 58.99                                 &                               &                               & 54.48                                  & 39.89                               & 66.85                                    & 46.06                                 \\
                                     & non-linear SVM                                        & 50.14                                             &                               &                               & 50.14                                  & 100.00                              & 0.00                                     & 66.79                                 &                               &                               & Null                                   & 0.00                                & 100.00                                   & 0.00                                  \\
                                     & ANN                                                   & 68.94                                             &                               &                               & 68.23                                  & 71.20                               & 66.67                                    & 69.68                                 &                               &                               & 69.71                                  & 66.67                               & 71.20                                    & 68.16                                 \\ \hline
\multirow{7}{*}{HOG}                 & LR                                                    & 67.57                                             &                               &                               & 67.57                                  & 67.93                               & 67.21                                    & 67.75                                 &                               &                               & 67.58                                  & 67.21                               & 67.93                                    & 67.40                                 \\
                                     & $k$NN                                                   & 66.21                                             &                               &                               & 66.30                                  & 66.30                               & 66.12                                    & 66.30                                 &                               &                               & 66.12                                  & 66.12                               & 66.30                                    & 66.12                                 \\
                                     & naive Bayesian                                        & 60.76                                             &                               &                               & 68.87                                  & 39.67                               & 81.97                                    & 50.34                                 &                               &                               & 57.47                                  & 81.97                               & 39.67                                    & 67.57                                 \\
                                     & RF                                                    & 64.31                                             &                               &                               & 62.33                                  & 72.83                               & 55.74                                    & 67.17                                 &                               &                               & 67.11                                  & 55.74                               & 72.83                                    & 60.90                                 \\
                                     & linear SVM                                            & 55.86                                             &                               &                               & 54.82                                  & 67.93                               & 43.72                                    & 60.68                                 &                               &                               & 57.55                                  & 43.72                               & 67.93                                    & 49.69                                 \\
                                     & non-linear SVM                                        & 50.14                                             &                               &                               & 50.14                                  & 100.00                              & 0.00                                     & 66.79                                 &                               &                               & Null                                   & 0.00                                & 100.00                                   & 0.00                                  \\
                                     & ANN                                                   & \textbf{76.02}                                    &                               &                               & 73.76                                  & 80.98                               & 71.04                                    & 77.20                                 &                               &                               & 78.79                                  & 71.04                               & 80.98                                    & 74.71                                 \\ \hline
\multirow{7}{*}{LBP}                 & LR                                                    & \textbf{65.67}                                    &                               &                               & 68.35                                  & 58.70                               & 72.68                                    & 63.16                                 &                               &                               & 63.64                                  & 72.68                               & 58.70                                    & 67.86                                 \\
                                     & $k$NN                                                   & 55.86                                             &                               &                               & 55.85                                  & 57.07                               & 54.64                                    & 56.45                                 &                               &                               & 55.87                                  & 54.64                               & 57.07                                    & 55.25                                 \\
                                     & naive Bayesian                                        & 53.95                                             &                               &                               & 54.39                                  & 50.54                               & 57.38                                    & 52.39                                 &                               &                               & 53.57                                  & 57.38                               & 50.54                                    & 55.41                                 \\
                                     & RF                                                    & 56.95                                             &                               &                               & 56.25                                  & 63.59                               & 50.27                                    & 59.69                                 &                               &                               & 57.86                                  & 50.27                               & 63.59                                    & 53.80                                 \\
                                     & linear SVM                                            & 48.50                                             &                               &                               & 48.54                                  & 45.11                               & 51.91                                    & 46.76                                 &                               &                               & 48.47                                  & 51.91                               & 45.11                                    & 50.13                                 \\
                                     & non-linear SVM                                        & 50.14                                             &                               &                               & 50.14                                  & 100.00                              & 0.00                                     & 66.79                                 &                               &                               & Null                                   & 0.00                                & 100.00                                   & 0.00                                  \\
                                     & ANN                                                   & 61.04                                             &                               &                               & 61.33                                  & 60.33                               & 61.75                                    & 60.82                                 &                               &                               & 60.75                                  & 61.75                               & 60.33                                    & 61.25                                 \\ \hline
\multirow{7}{*}{GLCM}                & LR                                                    & 62.67                                             &                               &                               & 67.41                                  & 49.46                               & 75.96                                    & 57.05                                 &                               &                               & 59.91                                  & 75.96                               & 49.46                                    & 66.99                                 \\
                                     & $k$NN                                                   & 64.03                                             &                               &                               & 63.68                                  & 65.76                               & 62.30                                    & 64.71                                 &                               &                               & 64.41                                  & 62.30                               & 65.76                                    & 63.33                                 \\
                                     & naive Bayesian                                        & 57.22                                             &                               &                               & 61.74                                  & 38.59                               & 75.96                                    & 47.49                                 &                               &                               & 55.16                                  & 75.96                               & 38.59                                    & 63.91                                 \\
                                     & RF                                                    & 60.76                                             &                               &                               & 59.71                                  & 66.85                               & 54.64                                    & 63.08                                 &                               &                               & 62.11                                  & 54.64                               & 66.85                                    & 58.14                                 \\
                                     & linear SVM                                            & 55.86                                             &                               &                               & 71.15                                  & 20.11                               & 91.80                                    & 31.36                                 &                               &                               & 53.33                                  & 91.80                               & 20.11                                    & 67.47                                 \\
                                     & non-linear SVM                                        & 56.95                                             &                               &                               & 66.25                                  & 28.80                               & 85.25                                    & 40.15                                 &                               &                               & 54.36                                  & 85.25                               & 28.80                                    & 66.38                                 \\
                                     & ANN                                                   & \textbf{65.67}                                    &                               &                               & 67.68                                  & 60.33                               & 71.04                                    & 63.79                                 &                               &                               & 64.04                                  & 71.04                               & 60.33                                    & 67.36                                 \\ \hline
\end{tabular}}
\end{table*}

According to Table. \ref{tbl2}, RF in EBHI has the best classification result on color histogram, reaching 73.57\%. $k$NN and ANN also have good classification effect, and the classification accuracy of all three of them reaches about 70\%. The accuracy of the other four classifiers are all distributed between 50\%-60\%, and the difference in effect with the above three classifiers is obvious, and the classification effect performs poorly.

The luminance histogram is another color feature that shows similar classification performance similar to color histogram since it is the mean of the three components of the image pixel color. RF shows robustness on both features and is still the best classifier reaching 72.75\%. The worst classifier has an accuracy of only 50\% or less.

Among the texture features, HOG features achieve the highest accuracy of 76.02\% using ANN. LR, $k$NN, naive Bayesian, and RF also achieve accuracy rates above 60\%. However, the two SVMs do not perform well on HOG, below 60\%.

LR classifier achieves the highest accuracy of 65.67\% using LBP features. After that ANN reaches the highest accuracy of 61.04\% and the SVM performs the least, even below 50\%.

On the four statistical values of GLCM, ANN achieves an accuracy of 65.67\%. The accuracy of the non-linear SVM classifier for the GLCM is 56.95\%, outperforming the linear SVM.

\subsection{Evaluation of Deep Learning based Classification Methods on EBHI}

In this paper, three popular CNN models and a novel ViT model are used to test the classification effectiveness on the EBHI dataset.  In this experiment, the training, validation, and test sets are partitioned in the ratio of 4:4:2. Each model is used with a learning rate of 0.00002, and the batch size is set to 32 for 100 epochs to observe the classification results of different models on EBHI. The experimental results are shown in Table. \ref{tbl3}.

\begin{table*}
\caption{Classification results of four deep learning classifiers on EBHI (In [\%]). The bold font in the table indicates the maximum value or the best index of the classification results of different categories.}\label{tbl3}
\resizebox{\textwidth}{20mm}{
\begin{tabular}{ccccclllll}
\hline
\multicolumn{1}{l}{\textbf{Model}} & \multicolumn{1}{l}{\textbf{Model size}} & \multicolumn{1}{l}{\textbf{best eopch}} & \multicolumn{1}{l}{\textbf{training time}} & \multicolumn{1}{l}{\textbf{Acc}} & \textbf{Category} & \textbf{Precision} & \textbf{Recall} & \textbf{Specificity} & \textbf{F1-score} \\ \hline
\multirow{2}{*}{VGG16}             & \multirow{2}{*}{512.24}                 & \multirow{2}{*}{100}                    & \multirow{2}{*}{2318}                      & \multirow{2}{*}{\textbf{95.37}}  & Malignant          & 96.4               & 94.3            & 96.5                 & 95.3              \\
                                   &                                         &                                         &                                            &                                  & Benign            & 94.4               & 96.5            & 94.3                 & 95.4              \\ \hline
\multirow{2}{*}{Inception-V3}      & \multirow{2}{*}{83.12}                  & \multirow{2}{*}{93}                     & \multirow{2}{*}{2068}                      & \multirow{2}{*}{72.93}           & Malignant          & 73.2               & 72.6            & 73.3                 & 72.9              \\
                                   &                                         &                                         &                                            &                                  & Benign            & 72.7               & 73.3            & 72.6                 & 73.0                \\ \hline
\multirow{2}{*}{ResNet50}          & \multirow{2}{*}{89.69}                  & \multirow{2}{*}{99}                     & \multirow{2}{*}{2054}                      & \multirow{2}{*}{83.81}           & Malignant          & 91.4               & 74.7            & 92.9                 & 82.2              \\
                                   &                                         &                                         &                                            &                                  & Benign            & 78.6               & 92.9            & 74.7                 & 85.2              \\ \hline
\multirow{2}{*}{ViT}               & \multirow{2}{*}{31.17}                  & \multirow{2}{*}{99}                     & \multirow{2}{*}{2083}                      & \multirow{2}{*}{71.42}           & Malignant          & 71.4               & 71.7            & 71.1                 & 71.5              \\
                                   &                                         &                                         &                                            &                                  & Benign            & 71.5               & 71.1            & 71.7                 & 71.3              \\ \hline
\end{tabular}}
\end{table*}

Overall, the deep learning models perform better than classical machine learning methods. The VGG16 model has an accuracy of over 95\%. However, it takes a long time to train and has a much larger model size than other deep learning models.
Inception-V3, ResNet50 and ViT model have similar training times and the ResNet50 achieves the highest classification accuracy of 83.81\% among them. The transformer-based ViT model achieves an accuracy of 71.42\%.

\subsection{Discussion}

This section compares the classification results on the EBHI dataset using different classifiers from Linear Regression to Visual Transformer. The performance of different classifiers with features are completely different. Since the classical machine learning methods have a strict theoretical basis and simple ideas, they perform well on some specific features and algorithms. However, deep learning methods are still superior than most of the classical machine learning methods.

\section{Conclusion and Futures Works}

In this paper, we develop a new publicly available Colorectal Histopathological Image Dataset called EBHI with four magnifications of 5532 images: 40$\times$, 100$\times$, 200$\times$ and 400$\times$, and five types images of tumor differentiation stages.
EBHI has the function to test whether the image classifier has high classification accuracy and good robustness. 
For classical machine learning methods, this paper focuses on the accuracy differences among classifiers, who perform at best at only 76\%, but at worst below 50\%.
For the deep learning methods, all four models perform excellent classification results, with the highest accuracy rate reaching over 95\%.
This paper focuses on the analysis of the four models in terms of accuracy, model size, training time, and other metrics. The classification experiments on EBHI show that the performance of the image classification experiments in this paper is sufficient to test the existing image classification methods.

\section*{Acknowledgements}

This work is supported by the "National Natural Science
Foundation of China" (No. 61806047) and the "Fundamental
Research Funds for the Central Universities" (No.
N2019003).We also thank Miss. Zixian Li and Mr. Guoxian
Li for their important discussion in this work.

\section*{Declaration of competing interest}
The authors declare that they have no conflict of interest in this paper.

\bibliographystyle{unsrt}
\bibliography{cas-refs}

\end{document}